\documentclass[12pt]{article}
\usepackage{amsmath}
\usepackage{graphicx, epsfig}
\usepackage{times}

\begin{document}
\begin{titlepage}
\title{Novel features in exclusive
 vector--meson production}
\author{M. A. Perfilov$^+$, S. M. Troshin$^*$, N. E. Tyurin$^*$\\[1ex]
\small \it $^+$ Physics Department,\\
\small \it Moscow State University,\\
\small \it Vorobiovy Gory , Moscow,  119 899 Russia\\
\small  \it $^*$ Institute for High Energy Physics,\\
\small  \it Protvino, Moscow Region, 142281, Russia}
\normalsize
\date{}
\maketitle

\begin{abstract}
It is shown that the universality of the
initial and final state interactions responsible for the transition between the
on-- and off--mass shell states leads to the
energy independence of the ratio   of exclusive
$\rho$ electroproduction cross section to the total cross section.
It is  demonstrated that the above universality and explicit mass
dependence of the exponent  in the power--like energy behavior of the
cross-section  obtained in the
approach based on unitarity is in a quantitative
agreement with the high--energy HERA experimental data.
We discuss also  HERA results on angular distributions
of vector--meson production.

\end{abstract}
\end{titlepage}
\setcounter{page}{2}

\section*{Introduction}
Exclusive vector meson production is an important process which can
provide information on hadronic structure at large and small distances
and nature of soft and hard interaction dynamics.
As it follows from the HERA data \cite{zosa,melld} the
 integral  cross section
of the elastic vector meson production $\sigma^{V}_{\gamma^* p}(W^2,Q^2)$
 increases with energy in a way
similar to the
$\sigma^{tot}_{\gamma^*p}(W^2, Q^2)$ dependence
 on  $W^2$ \cite{her}.
It appeared also that the growth of  the vector--meson
electroproduction cross--section  with energy is steeper for
heavier vector mesons. Similar effect takes also place when the virtuality
$Q^2$ increases.

The preliminary data of ZEUS Collaboration \cite{levy}
provide an indication for an
 energy independence of the
 ratio of the cross section of exclusive
$\rho$ electroproduction to the total cross section.
Such behavior of
this ratio is at variance with perturbative QCD results \cite{brod}, Regge and dipole
approaches \cite{levy,bial}. Recent review of the related   problems and successes
of the various theoretical approaches can be
found in \cite{land}.

Of course, the energy range of the available data is limited and
the above mentioned contradiction could probably be avoided due to fine tuning of
 the appropriate models.
Meanwhile, similar energy independence was  obtained in the approach
based on the off-shell extension of the $s$--channel unitarity \cite{epj02}.
In this note we provide an additional discussion pointing out to the origin of the above
energy independence. We  perform  a quantitative comparison of the HERA
data on vector--meson production with the results obtained in \cite{epj02}.

\section*{Vector--meson electroproduction}

There is no  universal, generally accepted
method to obey unitarity of the scattering matrix.
However, long time ago the arguments based on analytical properties
 of the scattering
amplitude  were put forward \cite{blan} in favor of the rational form of unitarization.
Unitarity  can be written for both
real and virtual external particles scattering amplitudes.
However, implications of unitarity  are different
 for the scattering of real and virtual
particles.
The extension of the $U$--matrix unitarization scheme
 (rational form of unitarization)
for the off-shell
scattering was considered in \cite{epj02}. It was supposed as usual that
the virtual  photon fluctuates into a quark--antiquark
 and this pair can
be treated as an effective virtual  vector meson state.
There were considered limitations the unitarity provides for the $\gamma^* p$--total
 cross-sections and geometrical effects in the
 energy dependence of $\sigma^{tot}_{\gamma^* p}$.
It was shown that the solution of the extended unitarity  augmented
by an assumption of
 the $Q^2$--dependent constituent quark\footnote{The concept of constituent quark has
 been  used extensively since the very beginning of the quark era but
  has obtained just recently a  possible
 direct experimental
 evidence at Jefferson Lab \cite{petron}.}
 interaction radius   results in the following
dependence at high energies:
\begin{equation}\label{sigt}
\sigma^{tot}_{\gamma^* p}\sim (W^2)^{\lambda(Q^2)},
\end{equation}
where
 $\lambda(Q^2)$ is saturated
 at large values of $Q^2$ and reaches unity. However, off--shell unitarity
 does not require transformation of this power-like dependence into a logarithmic
 one at asymptotical energies. Thus,  power--like behavior
 of the cross--sections
with the exponent dependent on virtuality
 could be of an asymptotical nature and have a  physical ground.
  It should not  be regarded merely as a transitory behavior
or a convenient way to represent the data.

The extended unitarity  for the off--mass--shell amplitudes $F^{**}$ and
$F^*$ has a structure  similar to the equation for the on--shell
amplitude $F$ but in the former case it relates the
different amplitudes. We denoted in that way the
amplitudes when both initial and final mesons  are off mass
shell, only initial meson is off mass shell and both mesons are on
mass shell, respectively. Note that $\sigma^{tot}_{\gamma^* p}$
 is determined by the imaginary part of the amplitude $F^{**}$,
  whereas $\sigma^{V}_{\gamma^* p}$ is
 determined by the square of  another amplitude $F^*$.
The important point in the solution of the extended unitarity
is the factorization in the impact parameter representation
at the level of the input dynamical
quantity~---~$U$--matrix:
\begin{equation}
U^{**}(s,b,Q^2)U(s,b)-[U^{*}(s,b,Q^2)]^2=0.\label{zr}
\end{equation}
Eq. (\ref{zr}) reflects universality of the initial and final state interactions
when transition between on-- and off--mass shell states occurs.
Despite that such factorization does not
survive  at the level of the
amplitudes $F^{**}(s,t,Q^2)$, $F^*(s,t,Q^2)$
and $F(s,t)$ (i.e.  after unitarity equations are solved and Fourier-Bessel
  transform is performed), it is essential for the energy independence of the
 ratio  of the exclusive
$\rho$ electroproduction cross section to the total cross section.

The above result (\ref{sigt}) is valid  when the interaction radius of the
constituent quark $Q$ from the virtual meson $V^*$ increases with
 virtuality  $Q^2$. We use the same notation $Q$ for the
constituent quarks and virtuality, but it should not lead to misunderstanding.
The dependence of the interaction radius
\begin{equation}\label{rqvi}
R_{Q}(Q^2)=\xi(Q^2)/m_Q.
\end{equation}
on  $Q^2$ comes through the dependence
of the universal $Q^2$-dependent factor $\xi(Q^2)$
(in the on-shell limit $\xi(Q^2)\to\xi$).
The origin of the rising
 interaction radius of the
constituent quark  $Q$ with virtuality $Q^2$
 might be of a dynamical nature and it would steam from
 the emission of the additional $q\bar q$--pairs in the
 nonperturbative  structure of a constituent quark.
Available experimental data
 are consistent with the $\ln Q^2$--dependence of the radius
 \cite{epj02}:
\[
R_Q(Q^2)=R_Q^0 + \frac{a}{m_Q}\ln\left(1+ {Q^2}/{Q_0^2}\right),
\]
where $R_Q^0=\xi/m_Q$ and parameters $\xi$, $a$ and $Q_0^2$ are
universal for all constituent quarks.

The introduction of the $Q^2$ dependent   interaction radius of a constituent
quark, which in this approach consists of a current quark
surrounded by the  cloud of quark--antiquark pairs of different
flavors \cite{csn}, is the main issue of the off--shell extension of the
model, which provides at large values of
$W^2$
\begin{equation}\label{totv}
\sigma^{tot}_{\gamma^* p}(W^2,Q^2)\propto G(Q^2)\left(\frac{W^2}{m_Q^2}
\right)^{\lambda (Q^2)}
\ln \frac{W^2}{m_Q^2},
\end{equation}
where
\begin{equation}\label{lamb}
\lambda(Q^2)=1-{\xi}/{\xi(Q^2)},\quad
\xi(Q^2)=\xi + a\ln\left(1+ {Q^2}/{Q_0^2}\right).
\end{equation}
 The value of parameter $\xi$
 in the model is determined by the slope of the differential cross--section
of elastic scattering at large $t$ region \cite{lang}
and it follows from the $pp$-experimental data  that $\xi=2$.

Inclusion of heavy vector meson production into this
scheme is straightforward: the virtual photon fluctuates before
the interaction with proton into the heavy quark--antiquark pair
 which constitutes
the virtual heavy vector meson state. After interaction with a proton
this state turns out into a real  heavy vector meson.

Integral exclusive (elastic) cross--section of vector meson production in
the process $\gamma^*p\to Vp$ when the vector meson in the final
state contains not necessarily  light quarks can be calculated directly:
\begin{equation}\label{elvec}
\sigma^{V}_{\gamma^* p}(W^2,Q^2)\propto G_{V}(Q^2)\left(\frac{W^2}
{{m_Q}^2}
\right)^{\lambda_{V} (Q^2)}
\ln \frac{W^2}{{m_Q}^2},
\end{equation}
where
\begin{equation}\label{lavm}
\lambda_{V}(Q^2)= \lambda (Q^2){\tilde{m}_Q}/{\langle m_Q \rangle}.
\end{equation}
In Eq. (\ref{lavm}) $\tilde{m}_Q$ denotes the mass of the constituent
quarks from
the vector meson and
$\langle m_Q \rangle =(2\tilde{m}_Q+3m_Q)/5$
 is the mean constituent
quark mass
of the vector meson and proton system.
Of course, for the on--shell scattering ($Q^2=0$) we have a standard Froissart--like
asymptotic energy dependence.

It is evident from Eqs. (\ref{totv}) and (\ref{elvec}) that
$\lambda_{V}(Q^2)=\lambda(Q^2)$
for the light vector mesons, i.e. the ratio
\begin{equation}\label{rv}
r_V(W^2,Q^2)=\sigma^{V}_{\gamma^* p}(W^2,Q^2)/\sigma^{tot}_{\gamma^* p}(W^2,Q^2)
\end{equation}
does not depend on energy for $V=\rho,\omega$.  Eq. (\ref{elvec}) and
consequently (\ref{rv}) are
in a good agreement (Fig. 1) with the experimental data of H1 and ZEUS Collaborations
\cite{zosa,melld}. It should be noted, however, that the experimental definition of the ratio
$r_V$ adopted by the ZEUS Collaboration \cite{levy} considers cross--sections
 at the different
 virtualities\footnote{We are indebted to I. Ivanov for pointing out to this fact.} and
 does not directly correspond to (\ref{rv}).
\begin{center}
\begin{figure}[h]
\begin{center}
\epsfxsize=2.6in \epsfysize= 2.2in
 \epsffile{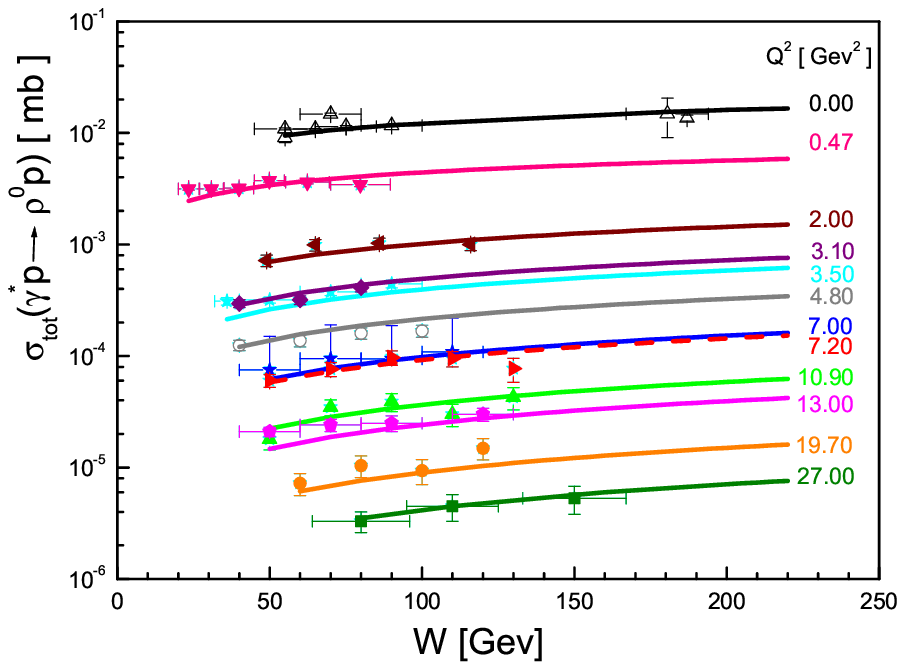}\;
\epsfxsize=2.6in \epsfysize= 2.2in
 \epsffile{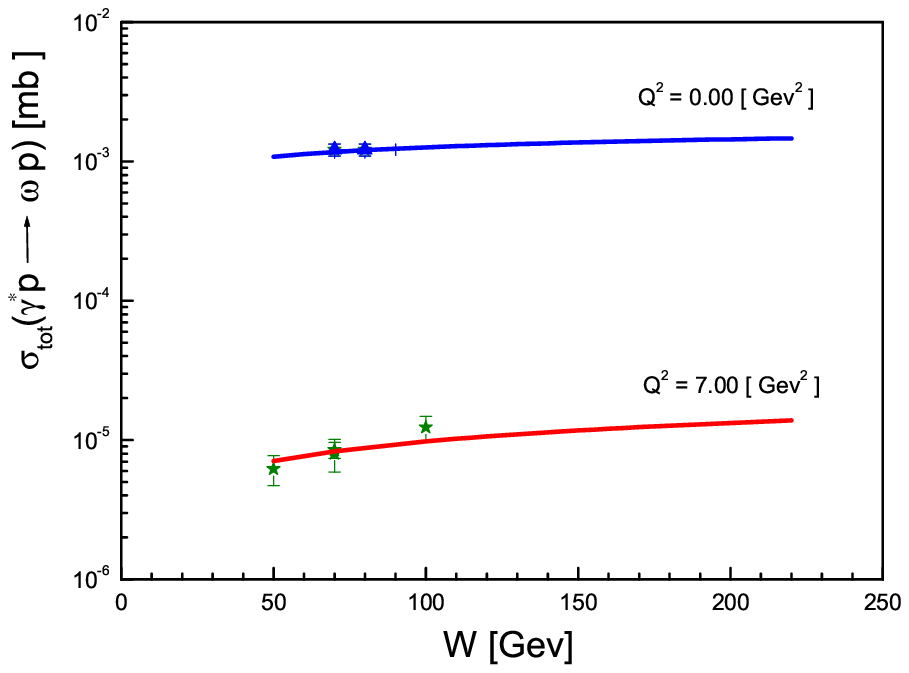}
\end{center}
 \caption[pdep]{Energy dependence of the elastic
 cross--sections of exclusive $\rho$ and $\omega$--meson production.}
\label{fig:1}
\end{figure}
\end{center}

For the case of the heavy vector meson production $J/\Psi$ and $\Upsilon $ the
 respective cross--section
rises about two times  faster than the total  cross--section;
Eq. (\ref{lavm}) results in
\[
\lambda_{J/\Psi}(Q^2)\simeq 2\lambda(Q^2),\;
\lambda_{\Upsilon}(Q^2)\simeq 2.2\lambda(Q^2),
\]
i.e.
\[
r_{J/\Psi}(W^2,Q^2) \propto (W^2)^{\lambda(Q^2)},\;
r_{\Upsilon }(W^2,Q^2) \propto (W^2)^{1.2\lambda(Q^2)}.
\]
Corresponding relations for the $\varphi$--meson production
 are the following
\[
\lambda_{\varphi}(Q^2)\simeq 1.3\lambda(Q^2),\;
r_{\varphi}(W^2,Q^2) \propto (W^2)^{0.3\lambda(Q^2)}.
\]
In the limiting case when the vector meson
is very heavy, i.e. $\tilde m_Q\gg m_Q$ the relation
between exponents is
\[
\lambda_{V}(Q^2)=2.5\lambda(Q^2).
\]
The quantitative agreement  of Eq. (\ref{elvec})  with experimental data for the case
of $\varphi$ and $J/\psi$ production can be seen in Fig. 2
\begin{center}
\begin{figure}[htb]
\begin{center}
\epsfxsize=2.6in \epsfysize= 2.1in
 \epsffile{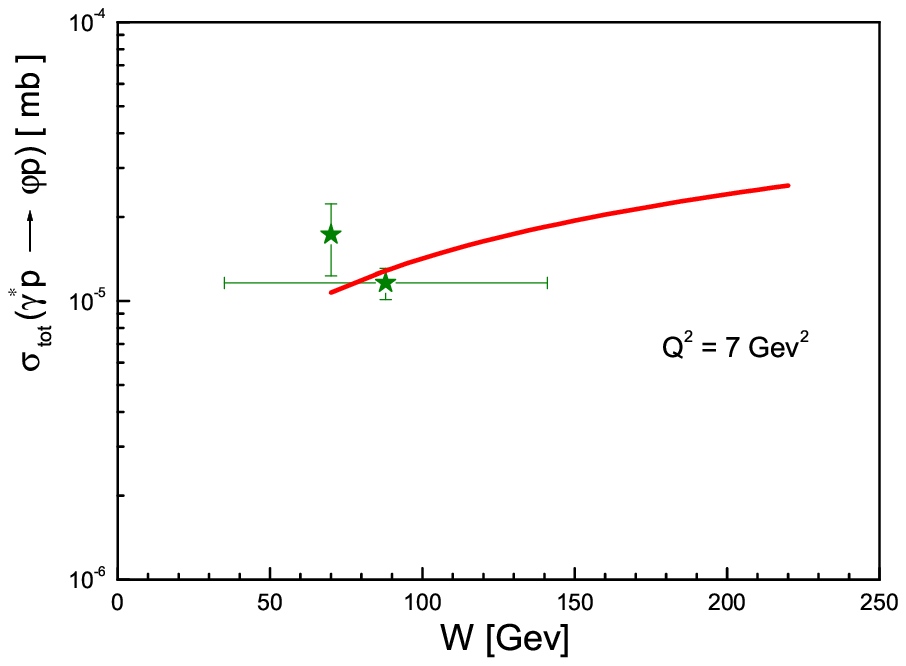}\;
\epsfxsize=2.6in \epsfysize= 2.1in
 \epsffile{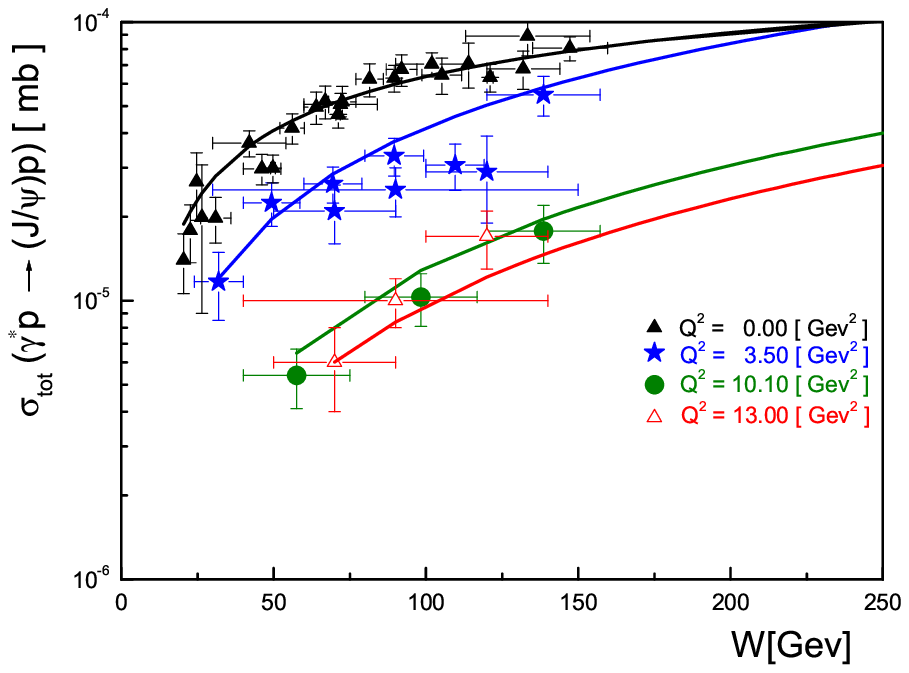}
\end{center}
 \caption[pdep]{Energy dependence of the elastic
 cross--sections of exclusive $\phi$ and $J/\psi$--meson production.}
\label{fig:2}
\end{figure}
\end{center}
This  agreement is in favor of relation (\ref{lavm}) which provides explicit
mass dependence of the exponent  in the power--like energy dependence of cross--sections.
Thus, the power-like parameterization of the
ratio $r_V$
\[
r_V(W^2,Q^2)\sim (W^2)^{\lambda (Q^2){(\tilde{m}_Q-\langle m_Q \rangle)}/
{\langle m_Q \rangle}}
\]
 with $m_Q$ and $Q^2$--dependent
exponent could also have a physical ground.  It would be interested therefore
to check experimentally the predicted energy dependence of the
ratio $r_V$.

The dependence of the constituent quark
interaction radius  (\ref{rqvi}) on its mass and virtuality
gets an experimental support and the
 non--universal energy asymptotical dependence (\ref{elvec}) and (\ref{lavm}) and
predicted  in \cite{epj02}
 does not  contradict to
the high--energy experimental data on elastic
vector--meson electroproduction. Of course, as it was already mentioned in the
Introduction, the limited energy range of the available experimental data  allows
other parameterizations, e. g.  universal asymptotical Regge--type behavior
with $Q^2$--independent trajectories (cf. \cite{petr,mart,fior}), to treat
the  experimental
 regularities as transitory ones.

It seems, however, that the scattering of virtual
particles reaches the  asymptotics much faster than the scattering of
the real particles and the $Q^2$--dependent exponent $\lambda(Q^2)$ reflects
the asymptotical
dependence and not the "effective" pre\-as\-ymp\-totical one.
Despite that the relation  between $\xi(Q^2)$ and $\lambda(Q^2)$
 implies  a saturation of the $Q^2$-dependence of $\lambda(Q^2)$ at large
values of $Q^2$, the power--like energy dependence itself will survive
 at asymptotical energy values.
The early asymptotics of virtual particle scattering
is correlated with the peripheral  impact parameter
 behavior of the scattering amplitude
for the  virtual particles. The respective profiles of the amplitudes $F^{**}$
and $F^{*}$ are peripheral when $\xi(Q^2)$
increases with $Q^2$ \cite{epj02}.

The energy independence of the ratio
 $r_\rho(W^2,Q^2)$
  reflects
 universality of the
initial and final state interactions responsible for the transition between the
on-- and off--mass shell states.
This universality is a quite  natural assumption
leading to factorization (\ref{zr}) at the level of the $U$--matrix \cite{epj02}.
Under this the  off--shell unitarity is the principal origin of the
energy independence of the ratio $r_\rho(W^2,Q^2)$.

There are also other interesting manifestations of the off--shell unitarity
effects, e.g. the behavior of the differential cross--sections at large $t$ is to a
large
extent determined by the off-shell unitarity effects. Indeed, a smooth power--like
 dependence
on $t$ has been predicted \cite{epj02}:
\begin{equation}\label{ttri}
  \frac{d\sigma_V}{dt}\simeq \tilde{G}(Q^2)\left[1-{\bar{\xi}^2(Q^2)t}/{\tilde{m}_Q^2}
  \right]^{-3},
\end{equation}
where
\begin{equation}\label{ksi}
  \bar{\xi}(Q^2)={\xi\xi(Q^2)}/[{\xi-\xi(Q^2)}].
\end{equation}
As it is seen from Fig. 3, Eq. (\ref{ttri}) corresponds to  experimental data for $J/\psi$
vector meson
even at the very moderate $t$-values. This fact is due to the effect
of large mass of charmed quark which enhances in the model contribution of the region
of  small impact
parameters.
\begin{center}
\begin{figure}[h]
\begin{center}
\epsfxsize=3in \epsfysize= 2.5in
 \epsffile{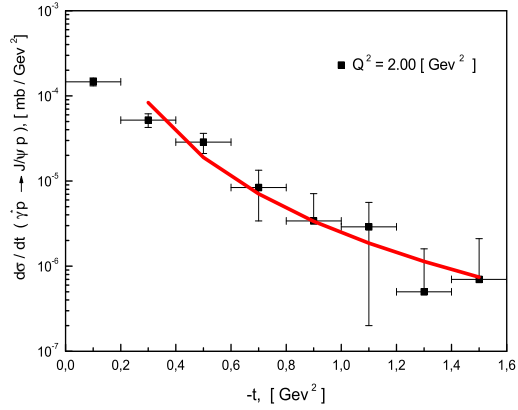}
\end{center}
 \caption[pdep]{Angular dependence of the elastic
 cross--sections of exclusive $J/\psi$--meson electroproduction.}
\label{fig:3}
\end{figure}
\end{center}
It appears that differential cross section does not depend  on energy and
depends on $t$ in a simple power-like way
$(-t)^{-3}$.
This dependence  differs from the corresponding dependence
in the case of  on-shell exclusive
scattering \cite{csn} which approximates the quark  counting rule \cite{matv}.
The ratio of differential cross-sections for the exclusive production of the
different vector mesons
\[\frac{d\sigma_{V_1}}{dt}/\frac{d\sigma_{V_2}}{dt}\]  does
not depend on the variables $W^2$ and $t$ at large enough values of $-t$.

Meanwhile the Orear type behavior of the differential cross-section of
the vector meson photoproduction obtained in \cite{epj02}
\begin{equation}\label{orev}
  \frac{d\sigma_V}{dt}\propto\exp\left[-\frac{2\pi\xi}{M}\sqrt{-t}\right].
\end{equation}
is in a good agreement with the HERA experimental data at moderate values of
$t$ (cf. Fig.4). Note, that the parameters $\xi$ and $M=2\tilde{m}_Q +3m_Q$
 are fixed.
\begin{center}
\begin{figure}[h]
\begin{center}
\epsfxsize= 1.6in \epsfysize= 1.6in
 \epsffile{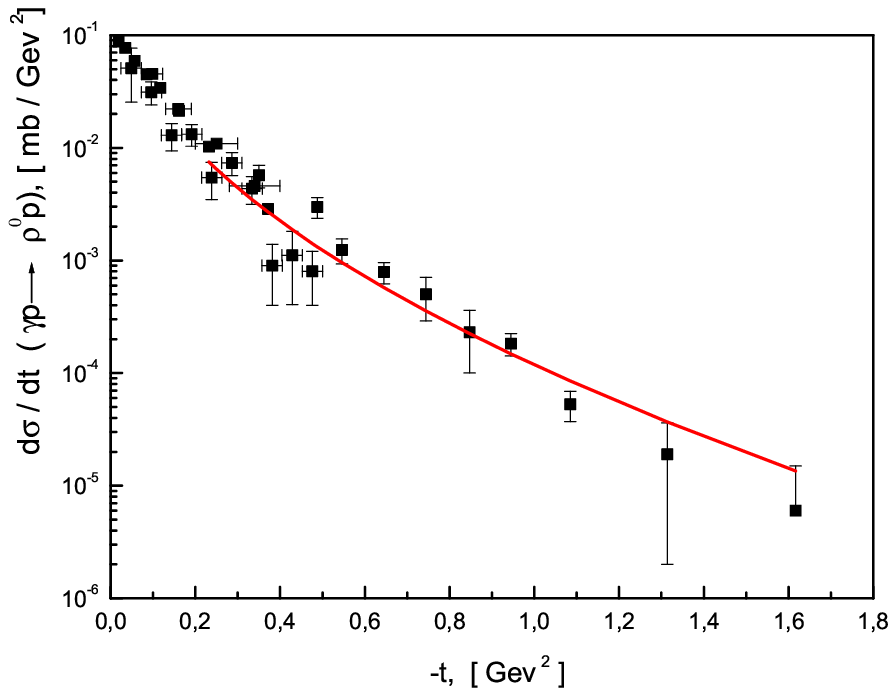}\quad
\epsfxsize=1.6in \epsfysize= 1.6in
 \epsffile{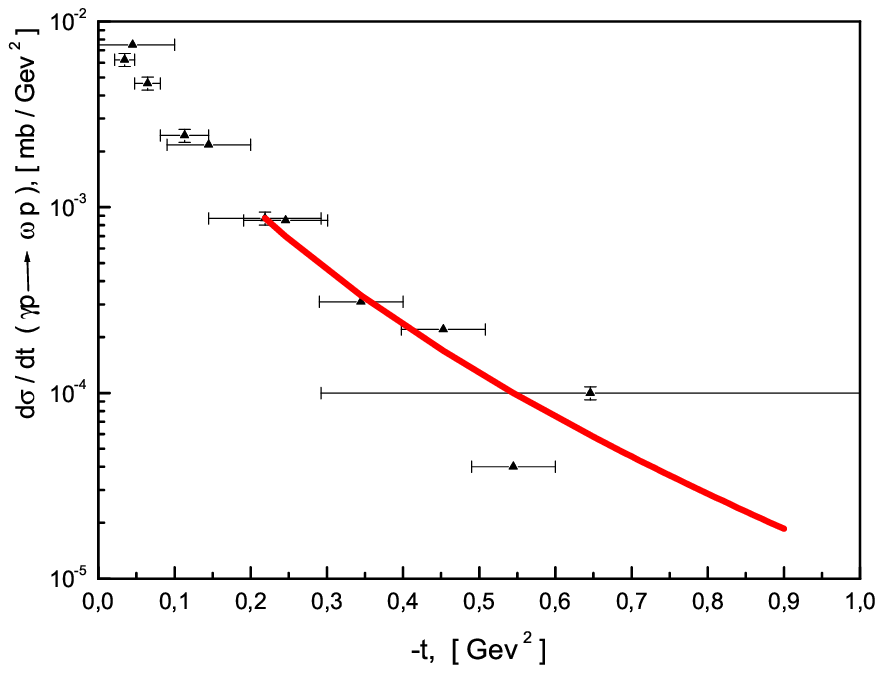}\quad
\epsfxsize=1.6in \epsfysize= 1.6in
 \epsffile{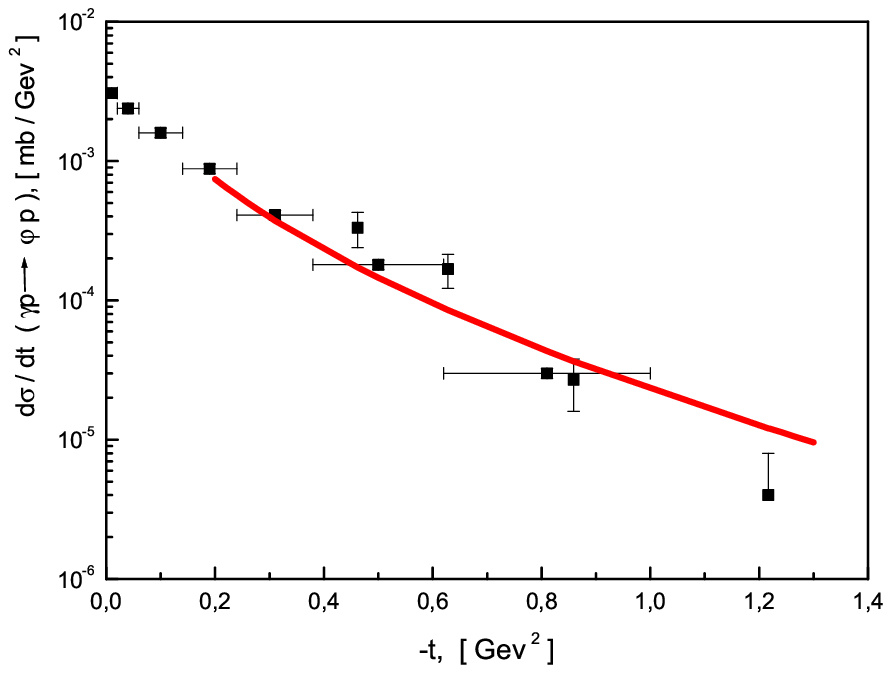}
\end{center}
 \caption[pdep]{Angular dependence of
  exclusive $\rho$, $\omega$ and $\varphi$--meson production.}
\label{fig:4}
\end{figure}
\end{center}
Thus, we have shown  that the model \cite{epj02}
which leads to the energy independence of the ratio of exclusive
$\rho$--meson electroproduction cross--section to the total cross--section as
the result of the adopted universality of the
initial and final state interactions responsible for the transition between the
on-- and off--mass shell states, is in
a quantitative agreement with experimental data
for the vector--meson
production.

\section*{Acknowledgement}
We are grateful to I.~Ivanov and V.~Petrov  for  the interesting
 discussions of the results and experimental situation.

\end{document}